# *Organizing and financing interstellar space projects – A bottom-up approach.*

*Dr. Frederik Ceyssens, drs. Maarten Driesen, drs. Kristof Wouters, dr. Pieter-Jan Ceyssens, drs. Lianggong Wen.*

*fceyssen@esat.kuleuven.ac.be*
*KULeuven, Belgium.*

Abstract

The development and deployment of interstellar missions will without doubt require orders of magnitude more resources than needed for current or past megaprojects (Apollo, Iter, LHC,…). Question is how enough resources for such gigaprojects can be found. In this contribution different scenarios will be explored assuming limited, moderate economic growth throughout the next centuries, i.e. without human population and productivity continuing to grow exponentially, and without extreme events such as economic collapse or singularity.

In such a world, which is not unlike the current situation, gigascale space projects face a combination of inhibiting factors: the enormous cost threshold, the need for risky and costly development of often quite application specific technology, the relatively little benefit with respect to the costs for the sponsors, the time span of at least a few generations and the absence of a sense of urgency.

The likelihood of direct commercial investment is found to be slim. The likelihood of governmental investment in several economic growth scenarios is assessed based on a range of political and economical factors behind successful and cancelled megaprojects, and found to be uncertain but possible.

Finally, it will be argued that the best chance of getting an interstellar project started in this generation is to establish an international network of non governmental organizations (NGOs) focused on private and public fundraising for interstellar exploration, similar to e.g. the WWF. It is shown that this path reduces the massive barriers to entry as well as the level of governmental support needed. The path is argued to fit best with three defining features of gigascale space projects, which should be recognized to the fullest: their almost absolute nonprofit character, their massive cost in terms of time and resources and their non-urgency leading to procrastination.

The NGO should set a single inspiring goal: to get a second home planet for humanity and the rest of Earth's life forms by the end of the millennium. Even if relatively modest but recurring funding is found enough resources can be gathered over time, provided the majority of the NGOs' income is saved up in a long term fund similar to that of the Noble Foundation or Norway's Government Pension Fund.

## Introduction

Besides paradigm-shifting scientific insights and technological advances produced by individuals or small teams, the previous century is also marked by over a dozen so-called megascience or 'big science' projects such as the Manhattan project, the space race, the Hubble Space Telescope and the Large Hadron Collider.

Such technoscientific projects are characterized by a high barrier to entry for even a single experiment or facility. Thus, their results could not have been realized incrementally in a traditional small scale research setting. The investment of over 1 billion US dollar of public money in each of these megaprojects could be justified by the high expected returns in terms of military supremacy, energy independence, support of the high tech industry or the appeal and prestige associated with a spectacular advance of technology and fundamental science [1-2].

Some have already contemplated on projects requiring an investment of resources that is even several orders of magnitude higher, in the trillions of dollars. These are often related to the expansion of humanity into space. Several more or less credible gigascale space projects have actually been proposed. Examples include projects on interstellar exploration [3-4], settlement on a habitable world yet to be discovered [5], or the terraforming[6-7] of a world in our own solar system. Many of these proposals are in principle feasible based on our current knowledge of physics at attainable energy levels. In contrast to popular opinion, no paradigm-shifting discoveries in fundamental physics are needed. Tempting (or as some state, quintessential for the long term survival and prosperity of humanity [8]) as they may be, clearly little is moving towards their actual realization.

What keeps these proposals from being more than a pipedream? It must be pointed out that the costs and time spans involved are not as unique as might seem: at this very moment mankind is successfully working on several trillion dollar size multigenerational enterprises, such as the buildup of the energy infrastructure or the ongoing development of integrated circuits. These cannot be seen as single gigascale projects though: there is no almost insurmountable threshold involved, but rather a chain of smaller steps each leading to commercially viable results along the way.

As elaborated further on, in the current state of affairs gigascale space projects face a combination of inhibiting factors. First of all, there is the enormous cost threshold. Furthermore, risky and costly basic technology development is still required and some of the technologies needed are quite application specific and unlikely to result from other areas of research. From a sociopolitical standpoint, even though promising a majestic future for mankind, these projects offer relatively little benefit with respect to the costs to the sponsors or investors themselves. Finally, the projects will necessarily span at least a few generations and no sense of urgency is present.

The trillion dollar question remains: given these factors, how can gigascale space exploration projects be organized and financed now or in the future? We will assume spacecraft as for example described in [3-4], based on our current knowledge of physics and technology realistically attainable in the near future such as nuclear fusion, implying a maximum speed around 0.1 c and many decades if not centuries of travel time. We will consider two cases: a relatively modest interstellar probe and a much larger (however in principle still feasible for an economy not much larger than the current) colonization ship.

Further assumptions

In most of this text, we will assume a moderate economic growth scenario in which both world population as well as productivity will not rise exponentially indefinitely, but will rather follow a sigmoid curve that reaches a relatively stable value in the future. Both world GDP and the general level of technology are still moving up. However, as elaborated below, it is far from certain that the 135-fold GDP increase seen over the last 3 centuries caused by the three industrial revolutions' rise in population (x11.1 in the period 1700-2008) and productivity (x12.3) [9] will be repeated.

An increased GDP is not likely to come from population increase: projections from the United Nations show population topping off around 10 billion in the second half of the 21$^{st}$ century under almost all scenarios [10-11]. As the average world GDP per person is still a factor 3 behind that of an average resident of Western Europe [9], more can certainly be gained in productivity. As a higher income at the moment still entails a higher use of nonrenewable resources and the carrying capacity of the planet is probably already surpassed (e.g. in terms of ecological footprint [12]), a transition to a sustainable resource and energy based economy will be required. If this is achieved there is no reason a relatively high level of wealth cannot be sustained, albeit no continued exponential growth will be feasible.

Another factor supporting the limited growth scenario is the fact that the rate of invention can be expected to continue to decline further [13]. Thus, we will assume there is a limit to the amount of wealth per person such an economy can produce and that world's economic output will eventually saturate in a sigmoid curve, perhaps a factor 2 to 5 above the current level. The world will be in a long lasting state of ecological and economical stability analogous to the island states that could be found on New Guinea or some Pacific islands such as Tonga or Tikopia, lasting millennia [14]. In terms of space development, technology will further evolve and allow economically realistic outposts for some tourism, science or mining. However, no massive colonization of the solar system will occur in the scenario, for the same reasons why no thriving nations are formed the Antarctic, the Gobi desert or the deep sea.

This scenario would still be considered optimistic by most. In a competing scenario, economic growth will go on exponentially for a very long time, possibly supported by a fourth industrial revolution or singularity [15-16]. We will assume this will not happen, or will not be allowed to happen. If it did, an exercise in how to organize and finance gigascale projects would become trivial, as their relative cost would soon shrink to that of a small scale science project today. Another competing scenario, severe shrinking or collapse of the economy is not interesting in the context of this paper for similar reasons.

Assumptions about the cost of the interstellar project

Although the conclusions drawn will be applicable to a wide range of interstellar exploration projects under the assumptions stated, it is instructive to have a scenario in mind. Therefore, we will assume a development and launch of one trillion current value US$ for a basic interstellar probe and 20 trillion for a first, relatively modest, interstellar colonization ship of as yet undetermined type (life seeder, sleeper ship, …) which could be called for if a suitable destination planet would be found. The assumed per weight cost is still rather small compared to current typical space missions. However, it must be emphasized that the ships will be assembled in space and that most of the weight will be fuel or, for a colonisation ship, other bulk materials and not high tech components. Also, due to the rise of commercial launch services, series production and upscaling launch costs to low Earth orbit can be safely expected to decline to under US$1000 per kg. A launch budget of 100 billion would then allow at least 100 kt of material for assembly of a starship to be launched from Earth, enough to cater for example for a Daedalus-like probe. A launch budget 15 times larger would allow a small colonization ship. These numbers even exclude the possibly more economical harvesting of materials in space.

## Commercial investment

Can an interstellar mission be a successful business venture under the assumptions described above? As with all past megascience projects, direct commercial investment in these gigascale ventures is unlikely. Even in the long term enough direct financial return is not to be expected when considering the establishment of a trade colony. All realistic (in terms defined above) interstellar missions proposed up to now are one way. Until a second foothold of civilization is established in the other star system, which can take centuries in addition to the travel and preparation time, only communication signals can be sent back. Then question would be what kind of trade would be established. Even if any credible means of generating revenue were found, the very large time spans involved would still exponentially enlarge the already enormous initial capital cost even assuming unrealistically low interest rates [17]. The great amounts of upfront investment needed are very unattractive even for the most incautious investors For example, due to amortization with a capital cost of a mere 1% per year, the project's profits should be a discount factor $1.01^{1000} = 20959$ times higher than the initial investment in order just to achieve breakeven.

A second business option would be to let settlers pay for their trip. However, in that case still at the very least a few centuries of preparatory technical and exploratory (scouting probes etc.) work would be required, without which only a handful of risk-taking and (necessarily) wealthy individuals would pay upfront. Thus, again the same amortization and risk related disincentives come into play.

Of course, this does not exclude a large involvement of the private sector as supplier for the projects, similar to the use of a commercial service to build or launch an interplanetary exploration probe. Also, this does not rule out certain commercial space ventures such as asteroid mining or tourism. Under the right technological and economic circumstances, they can grow into profitable gigascale enterprises in a step by step fashion investors prefer. In the process, they can bring about technology relevant for space exploration gigaprojects as well, such as cheaper access to low earth orbit. In contrast to commercial space ventures and although offering a great prospect to science, humanity or even the entirety of life on Earth, the interstellar exploration and colonization projects mentioned above cannot offer any direct financial potential to investors.

## Classical government sponsorship

Governments do not have to base spending on financial return. However, as stated above governments will require return in some form and will weigh such return against investment size and risk.

As clearly shown by for example the stark contrast between the 30B$ budget of the five leading space agencies combined and the annual military expenditures of over 1.5 trillion USD worldwide [18], a large international coalition could very well support space projects of unseen magnitude even today.

The Apollo program took off in a very favorable period. It could be very well supported because of its part in the space race of the cold war, the booming economy yielding surpluses and the general public excitement about new technology. Its funding level, both in terms of percentage of gross domestic product (0.4% of GDP) [19] and even in actualized money, has never been surpassed. This might be the limit governments are willing to spend on space adventure. Still, if a similar sense of urgency were present, this would mean a global effort could yield an annual investment of several hundreds of billions of dollars, dwarfing the Apollo project. Sadly, one of the defining characteristics of the interstellar space projects considered is their clear non-urgency. There are many good reasons to start investing in these projects. However, there are no real reasons to start doing so right now. Also, at least until an alien civilization is discovered, no strategic drivers are present for investment in interstellar projects driven by a military strategy.

The discovery of a possibly habitable planet in a nearby star system might spark enthusiasm for an internationally supported long term mission. However, any concrete proposal will still face stark criticism judging it to be unrealistic, too long term (crossing several generations), having too little benefit for the sponsors with respect to the huge cost and thus better being postponed indefinitely until superior and cheaper technology would presumably become available or some miraculous breakthrough in physics is achieved. As important parts of the needed technologies, especially those for propulsion as currently envisioned are very application specific this attitude could very well prevent them from ever being realized.

It is all too human that long term projects have a hard time gaining political capital when decision makers will not live to see the outcome. Even megaprojects promising huge returns in a couple of decades, such as the ITER fusion reactor, have taken over a decade to get approved. Space gigaprojects offer less hard return for the funding government, yet require at least an order of magnitude more resources and time. They will have a hard time competing with megaprojects promising shorter term and directly useful solutions for more urgent problems. Psychologically, future benefits for society seem to be discounted in a similar way as is done financially for long term investments discussed in the previous part.

In Table 1, the political decision factors behind several well known megaprojects are summarized, and compared with a colonization ship or interstellar probe mission. It is clear the latter do not compare favorably. Present day space enthusiasts can only wait and hope for a better future when technology, abundant resources, a stable political and societal climate, a long term vision and sense of adventure combine to foster enough political support for these high risk, ever low urgency, very long term projects. The probability estimation of such a future is left to the reader.

## A bottom-up path

However, a third path is possible that reduces the massive barriers to entry as well as the level of political support needed. This path starts with the thorough recognition of three distinguishing features of gigascale interstellar space projects: their almost absolute nonprofit character, their massive cost in terms of time and resources and their non-urgency leading to procrastination.

The large time span required and non urgency can actually be the key factor for the projects' acceptance: seen the long timespans required to reach the projects' goals, it is acceptable that the preparation and financing phase stretches out over generations as well. Because of the nonprofit character, it is not reasonable to ask a single generation to commit massive resources to the cause. However, as the projects are certainly worthwhile pursuing, it could be the privilege of each consecutive generation to contribute some intellectual and material resources. The construction of the great cathedrals in late medieval Europe can serve as an inspiring example.

Thus, in order to break the large financial, psychological and political barrier to entrance in practice, a long term fund similar to that of the Noble foundation or a government wealth fund can be established, fed with regular donations. Projections show that, in a reasonable range of expected financial returns and donations, a time span between one and five centuries will be needed to gather sufficient capital to back a gigascale project (figure 1). The fact that the technology required for the projects to succeed is not available yet must not be seen as a reason to delay establishing such a fund. Reversely, the very existence of the fund will act as an impetus to develop the technology, and even to attract funding for this from the usual sources. Knowing that realizing the projects will be tough but possible according to the laws of physics and that they will in principle be realistic seen the world's economic system is enough at the start. This way, the stalemate of waiting for dedicated technology to arrive can be broken. Both an interstellar probe as well as a colonization ship project fit well in this concept.

By building up the required capital bit by bit, it is no longer an insurmountable threshold. For example, if all major space agencies could be convinced to invest 10% of their budget in the fund (e.g. in return for a controlling stake in the organization), about 3 billion dollars would be fed into the fund yearly. Assuming a yearly return of 1.7% above inflation, the same average as the Noble Foundation achieved over the past 110 years [20], it would take 110 years to reach a target capital of 1000 billion (figure 1). This should be enough to back up a first interstellar mission, i.e. the technology development that remains to be done by the time the fund has reached its financial goal and the construction, launch and follow up of the project. In order to finance a colonization ship, little over 280 years would be required.

The concept could even work when relatively small yearly amounts, comparable to the revenue of a larger-sized nongovernmental organization (NGO) such as Greenpeace or the WWF, still several hundreds of millions of dollars, are fed into the fund. Then, it will be no longer necessary to convince the majority of political decision makers of the necessity of the project. A fairly sized gigascale project can even be realized by the continuous support of at the very least several hundreds of thousands of interested private supporters. Timespans are longer but not linearly: with a yearly investment of 300 million, the target of one trillion is reached in 245 years under the same conditions as used above.

The primary assumption relied on is that a significant part of the world retains a more or less liberal market economy during the next few centuries, and that this would allow attaining a long term capital gain higher than inflation with a well balanced, well spread and reasonably safe investment portfolio. This has been shown to be possible over a period of at least a century. As already mentioned, the tax-exempt Nobel Foundation has achieved an average return on investment of 1.7% above inflation despite world wars and economic crises. Tax exempt status must almost certainly be achieved in order for the project to succeed. The fund would become a major player in the world economy but it would not be disturbingly large. Even now, over 20 trillion dollars are gathered in pension funds worldwide [21].

## A possible scenario

To summarize the previous, for worthwhile long-term projects that are definitely well above normal big science budgets, the quest for financing should begin long before all technology is even developed. Knowing that they are realistic according to the laws of physics and that they can in principle be built by the global economy suffices. When the gradual buildup of a fund to finance the endeavor is taking momentum, this fact alone will encourage research and investment in its support. Thus, there seems to be no point in postponing. Now is as good a time as any.

As at the start probably not enough political and scientific capital will be present for major space agencies to back up an interstellar fund seriously. Therefore the second option, the establishment of an international NGO (juridically typically a network of national organizations) is the most likely to succeed. In contrast to existing space agencies, the NGO should have a single, clear and well marketable goal: to get a second home planet for humanity by the end of this millennium.

Existing space advocacy organizations such as the Planetary Society or the British Interplanetary Society could play a central role in establishing the initiative, and gain increased momentum. Other organizers could be interested individuals, academic and other institutions, etc.

We are convinced the lure of the vastness of space is strong enough to attract enough support for this effort, a level similar to that of existing worldwide operating NGOs. If it does not despite a serious effort of the organizers, this can be seen as a democratic signal of humanity that it is currently not interested in outer space exploration or colonization.

The NGO should, at least until governments step in, control the interstellar fund and put the majority of its income in it. Its other activities should be fundraising, lobbying and educating the public about realistic space travel. In order not to jeopardize its goal, it should operate as lean as possible.

When the NGO has gained momentum, it can decide to support research projects related to its goal however this should remain secondary to the buildup of the fund until a late stage. The major research interest should be into propulsion technology. The lesson of NERVA, of which the development was cancelled in order to make a costly manned Mars mission impossible [22], should be remembered. Once an adequate propulsion system is ready, popular and political support can be expected to increase significantly.

Conclusion

To conclude, gigascale space projects could bring about a majestic future for humanity but are hard to finance as they will span centuries, have a high financial barrier and do not alleviate any pressing need. However, even when no 100-fold growth of the global economy occurs, no profits are to be expected, no international consensus to shift a fair amount of resources to such enterprises arises and despite our natural preoccupation with the order of the day, they can come true if a long-term organization such as described above is established, manages to acquire reasonable levels of support and gradually builds up the required capital in a fund.

The fact that the required technology is not available yet must not be seen as a reason to delay building this fund. Reversely, the very existence of the fund will act as an impetus to develop the technology. Most likely, the organization will start as a non governmental organization or a network of such NGOs, similar to established worldwide charities such as the WWF.

Our generation will not see spacecraft depart for another star system. However, it could be our privilege to be able to lay the foundation of a something of unfathomable proportions.

| Proposed megaproject | Appoxim cost [G$] | Prospects | Consequence of non-action | remarks |
|---|---|---|---|---|
| Manhattan project | 22 | Decisive military advantage | Risk of nuclear Armageddon on own soil | completed |
| Apollo project | 100 | Cold war prestige, helps in arms race, technology development | Falling behind competitor nation | Succes, not fully completed |
| ITER | 20 | Energy independence, Clean sustainable economy, prestige | Risk of energy crisis | In progress |
| SSC particle collider | 12 | Fundamental discoveries in physics, prestige | Slower development of physics (non-urgent) | Cancelled during Construction (cost overrun) |
| Interstellar probe<br>Modest colonization ship | 1000?<br>20000? | Discovery, prestige, inspiration. Colony in other system, (BUT independent, takes 500+ years) | Impossibility of repopulating Earth after extinction event (non-urgent, in the far future) | Too much time & cost to be considered |

Table 1: Table summarizing some important decision factors behind the political decision making on past and current megaprojects and their comparison with possible future gigaprojects related to interstellar exploration. Costs are in USD actualized to 2008.

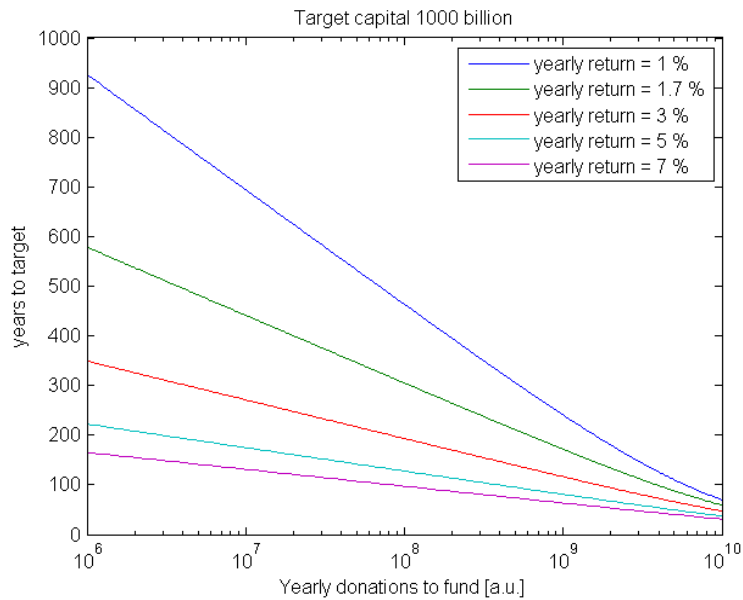

*Figure 1: Projected time until reaching a target of 1 trillion assuming a constant yearly return. Returns are after subtracting costs of investment and inflation.*

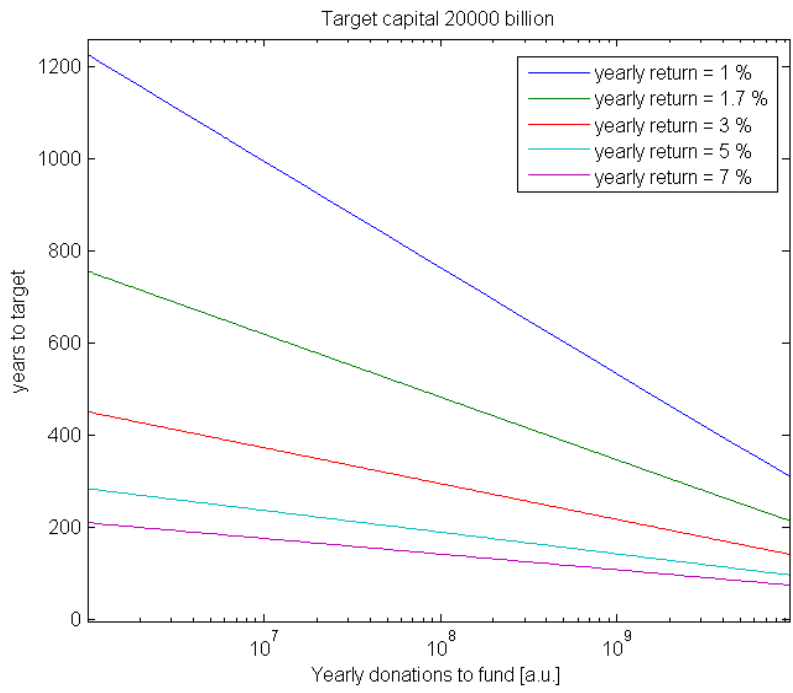

*Figure 2: Projected time until reaching a target of 20 trillion, same assumptions as in Figure 1.*